# Alternating superconducting and charge density wave monolayers within bulk 6R-TaS$_2$

A. Achari[1,2], J. Bekaert[3,4], V. Sreepal[1,2], A. Orekhov[4,5], P. Kumaravadivel[1,6], M. Kim[6], N. Gauquelin[4,5], P. Balakrishna Pillai[1,2], J. Verbeeck[4,5], F. M. Peeters[3], A. K. Geim[1,6], M. V. Milošević[3,4], R. R. Nair[1,2]

[1]National Graphene Institute, University of Manchester, Manchester, M13 9PL, UK.
[2]Department of Chemical Engineering, University of Manchester, Manchester, M13 9PL, UK.
[3]Department of Physics, University of Antwerp, Groenenborgerlaan 171, B-2020, Antwerp, Belgium.
[4]NANOlab Center of Excellence, University of Antwerp, Groenenborgerlaan 171, B-2020 Antwerp, Belgium.
[5]Electron Microscopy for Materials Science (EMAT), University of Antwerp, Groenenborgerlaan 171, B-2020 Antwerp, Belgium.
[6]Department of Physics and Astronomy, University of Manchester, Manchester, M13 9PL, UK.

**Van der Waals (vdW) heterostructures continue to attract intense interest as a route of designing materials with novel properties that cannot be found in naturally occurring materials. Unfortunately, this approach is currently limited to only a few layers that can be stacked on top of each other. Here we report a bulk material consisting of superconducting monolayers interlayered with monolayers displaying charge density waves (CDW). This bulk vdW heterostructure is created by phase transition of 1T-TaS$_2$ to 6R at 800 °C in an inert atmosphere. Electron microscopy analysis directly shows the presence of alternating 1T and 1H monolayers within the resulting bulk 6R phase. Its superconducting transition ($T_c$) is found at 2.6 K, exceeding the $T_c$ of the bulk 2H phase of TaS$_2$. The superconducting temperature can be further increased to 3.6 K by exfoliating 6R-TaS$_2$ and then restacking its layers. Using first-principles calculations, we argue that the coexistence of superconductivity and CDW within 6R-TaS$_2$ stems from amalgamation of the properties of adjacent 1H and 1T monolayers, where the former dominates the superconducting state and the latter the CDW behavior.**

Designing heterostructured materials with tailor-made properties has significant importance in fundamental and applied research. For example, the development of III-V semiconductor heterostructures[1] has transformed many aspects of our lives. More recently, the fabrication of heterostructures using two-dimensional (2D) materials with complementary properties[2] opens an astounding number of opportunities for designing exotic materials. Since 2D materials come in a plethora of different physical, chemical and electronic properties, the number of possible combinations we can achieve is unlimited, paving the way for materials with tailor-made properties. The layers in 2D heterostructures are held by their van der Waals interaction and are commonly referred to as van der Waals (vdW) heterostructures. They have already shown promise in different applications. For example, vdW heterostructures made using metal-insulator, metal-semiconductor and insulator-insulator 2D materials not only exhibit new physics (e.g., Hofstader butterfly states in graphene/hBN,[3] ultrafast charge transfer in $MoS_2$/$WS_2$ interface[4]) but show good performance in electronic and optoelectric applications such as field-effect transistors[5-7], photodetectors[8-10] and light-emitting diodes[11, 12]. In addition, some interesting properties of transition metal dichalcogenides such as 2D superconductivity[13] and charge density wave states[14] can also be tuned through interlayer coupling via vdW heterostructures.

Most of the vdW heterostructures are currently prepared by mechanically stacking one 2D layer on another. Even though this process is laborious, the precision in creating the heterostructure makes this process ideal for probing the fundamental properties of the heterostructure devices. On the other hand, the direct synthesis of vdW heterostructures in the chemical vapour deposition (CVD) process or via chemical methods is far from perfect. It is noteworthy that vdW heterostructure in bulk also exists in nature. Frankencite is a natural vdW heterostructure formed by alternate stacking of $SnS_2$ and $PbS$ layers[15]. Examples of such bulk vdW heterostructures are rare, and even with the progress in the 2D materials research, synthesising such bulk vdW heterostructures are still challenging. 6R phase of $TaS_2$ with alternating layers of 1H (superconducting)[16, 17] and 1T $TaS_2$ (Mott insulator)[18] is another example of bulk vdW heterostructure. This phase has rarely been studied due to the difficulty and inconsistency in synthesising the pure phase via the traditional vapour transport method[19, 20]. In this work, we report interlayered monolayer superconductivity and charge density waves (CDW) in 6R $TaS_2$ obtained by a thermally driven phase conversion of 1T $TaS_2$. We establish the bulk heterostructure of 1H and 1T layers in 6R phase by electron microscopy. We also show that superconductivity and CDW co-exist in this bulk vdW heterostructure, and the superconducting transition temperature (2.6 K) is higher than that in both 2H (0.8 K)[16] and 1T phases of TaS$_2$ (1.5 K at 2.5 GPa)[18]. Through exfoliation and restacking, we have demonstrated that the superconducting transition temperature can be further increased to 3.6 K by exfoliation and restacking of layers in 6R TaS$_2$.

Figure 1a shows the *in-situ* temperature-dependent evolution of the (001) XRD peak of a single crystalline 1T TaS$_2$ under a high vacuum. At temperatures below 500 °C, we only observed a peak shift associated with the thermal expansion of the crystalline *c* axis (Figure S1). But at 600 °C we noticed the appearance of a new peak at lower 2θ (14.78°) compared to the original peak of the 1T phase at 15.05°. Further increase in temperature resulted in steady decrease in 1T peak intensity with a concurrent increase in the intensity of the new peak at 14.78°. At 800 °C, no trace of the 1T peak remained in the XRD pattern. Since the sample was single-crystalline and highly oriented in nature, we only observed (*00l*) peaks in the XRD pattern. To fully understand the structural transformation after annealing, we performed further powder X-ray diffraction (PXRD) studies after grinding the 800 °C annealed crystal in a mortar. The resulting PXRD pattern shows peaks associated with other lattice planes in addition to (*00l*) peaks (Figure 1b). It is to be noted that the intensity of the peaks associated with the (*00l*) planes is higher due to the incomplete grinding. Upon matching the diffraction pattern to that of powdered 1T and simulated PXRD patterns from CIF files of 1T, 2H and 6R TaS$_2$, we found that it matches exclusively with 6R phase[21] (Figure S2). Figure 1b inset shows the zoomed view of the low angle peak and its comparison to the corresponding calculated peak positions of parent 1T [22] (blue line), 2H [23] (green line), and 6R (red line) phases of TaS$_2$. This further confirms that the 1T TaS$_2$ samples undergo phase transition while annealing in a vacuum. We have also performed control experiments where single-crystalline 2H TaS$_2$ was heated up to 800 °C. *In-situ* XRD does not show any phase change (Figure S3), suggesting only 1T phase can be converted into 6R phase by annealing.

The lower angle (006) X-ray peak in 6R TaS$_2$ suggest a lattice expansion in the *c*-direction after the phase change. The interlayer spacing corresponding to the (006) peak estimated from the XRD pattern was 0.597±0.001 nm for the 6R phase. On the other hand, the estimated interlayer spacing of the parent 1T phase was 0.590 nm, marking a 1.4% lattice expansion along the *c* direction.

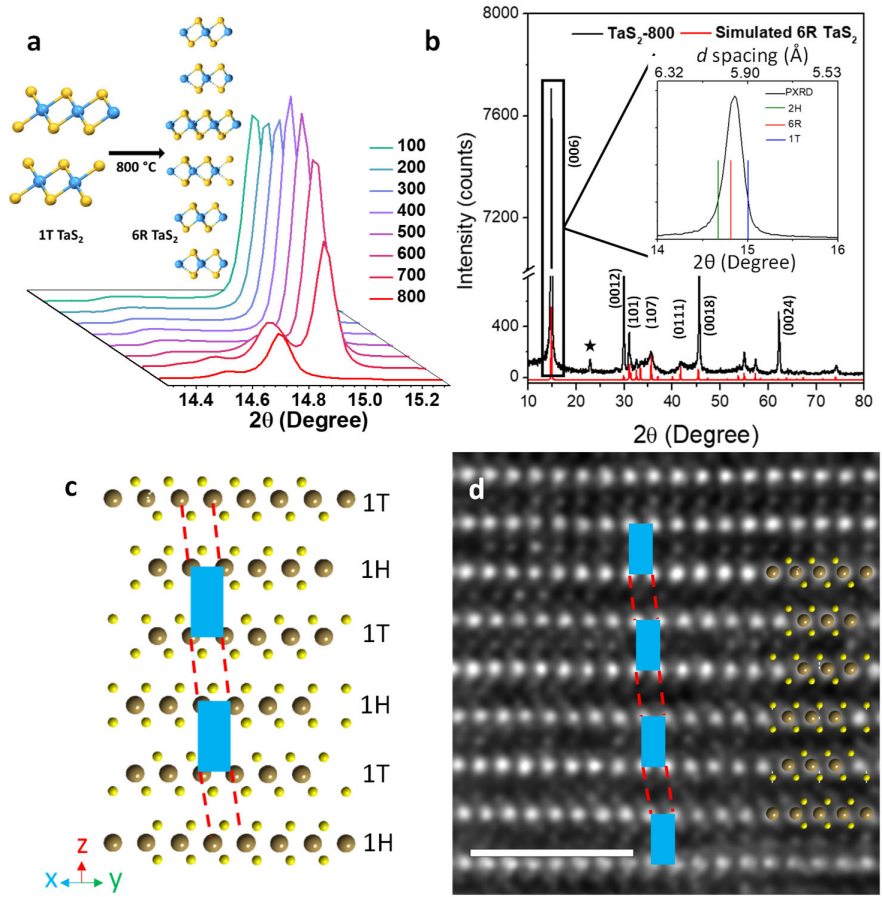

*Figure 1. 1T to 6R phase transition of TaS$_2$.* (a) In-situ temperature-dependent XRD of a 1T TaS$_2$ single-crystal heated in vacuum up to 800 °C. Inset: Schematic of phase transition of 1T TaS$_2$ to 6R phase. The teal spheres represent Ta and the yellow spheres represent S atoms. (b) PXRD pattern of the powdered form of 800 °C heated 1T crystal (black) compared to 6R phase reference spectra (red) by using the model[21]. * indicates peak from surface oxidation due to the residual air in the vacuum chamber. Inset: Zoomed view of the (006) peak shown inside the rectangle. Green, red, and blue lines show the (00l) peak position corresponding to 2H, 6R, and 1T phases of TaS$_2$, respectively. (c) model crystal structure of 6R TaS$_2$ showing alternating layers of 1T and 1H TaS$_2$. Blue rectangle and red dotted lines show each 1T-1H hetero layer are slightly displaced in the c-axis. (d) Cross-sectional high-resolution STEM image of annealed TaS$_2$ sample along [110] direction showing the alternating arrangement of 1H and 1T layers. Scale bar, 2nm. Overlaying 6R atomic model structure shows match of atomic positions and lattice stacking with the STEM image. In the model, Ta atoms are denoted as brown and S atoms as yellow spheres. The blue rectangle and the red dotted lines show that, similar to the model structure, each 1T-1H hetero layer is slightly displaced in the c-axis.

We have performed cross-sectional high-resolution scanning transmission electron microscopy (HRSTEM) of the annealed sample to confirm the phase transition into the 6R phase. Figure 1d shows a STEM image of the annealed sample at room temperature showing alternate stacks of 1T and 1H layers. We found a perfect match of atomic positions and lattice stacking of the annealed sample with a model 6R structure, ruling out the presence of any other polytype of $TaS_2$. Further, the interlayer spacing estimated from the intensity profiles (Figure S4) for the 1T to 6R phase (before and after heating) showed a 1.6% increase from 0.596±0.006 nm to 0.606±0.006 nm (Figure S5), indicating a transition to the 6R phase and closely matching the PXRD value. Additionally, the Raman spectra of bulk 6R sample shows presence of Raman active modes from both 1T and 2H phases, (Figure S6) confirming its heterobilayer structure.

Subsequently, we have investigated the electrical and magnetic properties of bulk 6R $TaS_2$. Figure 2a shows typical magnetisation vs temperature curves, $M(T)$, for 6R $TaS_2$ and parent 1T $TaS_2$ crystal under the external magnetic field of 5 Oe. Zero field cooling (ZFC) data for 6R $TaS_2$ clearly shows a diamagnetic transition at ≈2.5 K (shielding of the external field, $H$, which is characteristic of superconducting materials). The onset temperature of this superconducting transition is much higher than the transition temperature ($T_c$) of 2H phase (0.8 K). In comparison, the parent 1T phase or 1T $TaS_2$ annealed at different temperatures up to 600 °C does not show any diamagnetic transition as expected (Figure 2a and Figure S7). We have also carried out magnetic measurements on 2H $TaS_2$, and 2H $TaS_2$ heated at 800 °C, but no transition down to 1.8 K was observed (Figure S8), ruling out the presence of impurities or defects as a cause of the superconducting transition in the heated 1T sample. FC-ZFC measurements performed at higher fields reveal that the onset temperature of superconductivity ($T_c$) decreases when $H$ is increased (Figure 2a inset). No onset of superconductivity was observed at fields higher than 500 Oe. Upon exceeding this field, the samples show only a weak paramagnetic signal. The $M$-$H$ curve in Figure 2b exhibits a typical magnetic hysteresis profile of a type II superconductor. The phase diagram (Figure 2b inset) was obtained by calculating $H_{C2}$ at different temperatures from the divergence point on the X-axis. From the phase diagram, the $T_c$ of the material was estimated to be 2.6 K. The obtained 6R $TaS_2$ was stable in air. No change in magnetic measurements was observed after exposing the sample to an ambient atmosphere for one month.

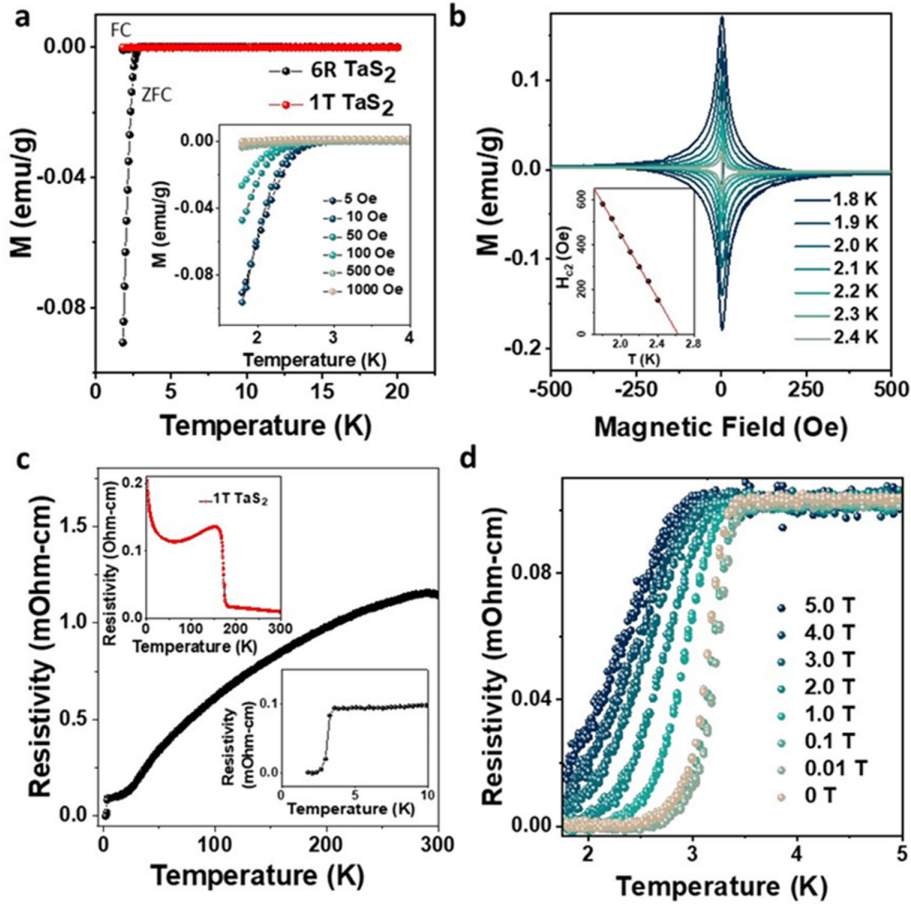

*Figure 2. Superconductivity in 6R TaS$_2$.* *(a) Temperature dependence of ZFC and FC magnetisation, M, for single-crystalline 6R TaS$_2$ and 1T TaS$_2$ under the magnetic field of 5 Oe applied parallel to the c axis. The inset shows ZFC and FC M(T) at different magnetic fields for 6R TaS$_2$. (b) Magnetisation dependence as a function of H‖c at different temperatures. The inset shows the temperature dependence of the upper critical field H$_{C2}$. The upper critical field was calculated from the divergence point in the M-H hysteresis curve. The solid red line is the guide to the eye. (c) Temperature dependence of electrical resistivity of 6R TaS$_2$ crystal at H = 0 T. Bottom inset shows zoomed superconducting transition. The top inset shows the temperature dependence of electrical resistivity of 1T TaS$_2$ nanosheets at H = 0 T. (d) The evolution of R(T) for 6R TaS$_2$ with increasing external magnetic field in an H‖ab geometry.*

TaS$_2$ being a layered material, we expect its superconductivity to be anisotropic across different crystalline directions. To investigate this, we studied the dependence of magnetic field orientation on the observed transition. All the magnetization studies so far were performed with a geometry where the *c*-axis of the TaS$_2$ crystal lies parallel to the magnetic field orientation ($H \parallel c$). Upon rotating the crystal plane orientation such that $H \parallel ab$, the observed diamagnetic transition is almost negligible (Figure S9a). Further, we have calculated the temperature dependence of critical magnetic field ($H_{C2}$) for parallel and perpendicular geometries and found the magnetic anisotropy ($H_{C2}^{\perp}/H_{C2}^{\parallel}$) to be 40 (Figure S9b). Such a high anisotropy indicates the 2D nature of superconductivity in 6R TaS$_2$. This value closely resembles the anisotropy observed in intercalated 2H TaS$_2$ (47),[24] and is much higher than other TaS$_2$ based systems such as 2H TaS$_2$ (6.7),[24] Pb$_{1/3}$TaS$_2$ (17),[25] restacked TaS$_2$ (11)[26] and 4Hb TaS$_2$ (17)[27].

Further evidence for superconductivity in 6R TaS$_2$ crystal was obtained from low-temperature electrical resistivity measurements. Figure 2c shows the zero-field resistivity of 6R TaS$_2$ plotted against temperature for the current flowing in the *ab* plane. At high temperatures, the resistivity decreases almost linearly with temperature showing the semi-metallic nature of 6R TaS$_2$. Upon lowering the temperature, a superconducting transition (where resistivity reaches zero) is observed at 2.6 K (Fig 2c bottom inset) in agreement with our magnetic measurements. On the other hand, the parent 1T TaS$_2$ (Figure 2c top inset) does not show any superconducting transition, but a large charge density wave (CDW) transition at 180 K from nearly commensurate CDW to commensurate CDW was observed as reported previously[28, 29]. Field dependent resistivity measurements on 6R TaS$_2$ were performed with the magnetic field parallel to the *ab* plane of the sample ($H \parallel ab$) and a decrease in superconducting transition temperature was observed with increasing field (Figure 2d).

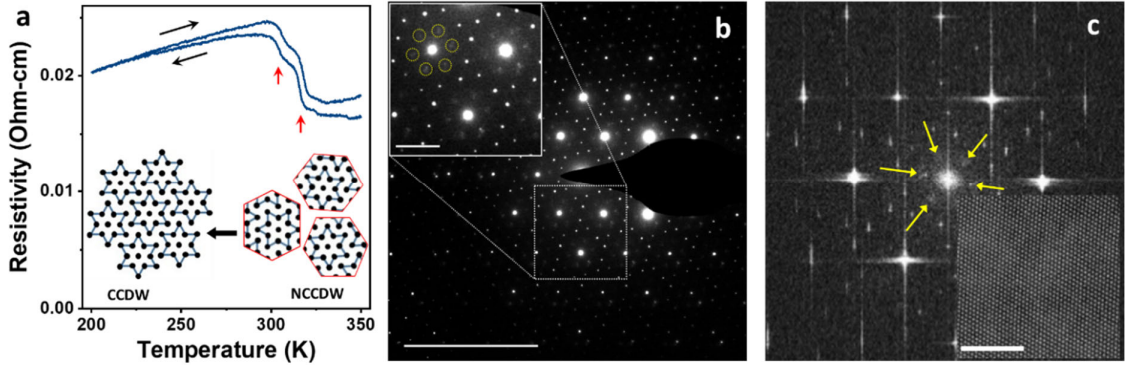

*Figure 3. Charge density wave (CDW) in 6R TaS$_2$. (a) Temperature-dependence of electrical resistivity of 6R TaS$_2$ showing CDW transitions (red arrows). Black arrows denote temperature sweep direction. Schematic representation of phase transition from nearly commensurate to commensurate structure is shown below. (b) Selected area electron diffraction pattern from 6R TaS$_2$ at room temperature (293 K). Scale bar, 10 1/nm. Top inset: zoomed view of the electron diffraction pattern from the white square marked area clearly showing CDW spots (Yellow circles). Scale bar, 2 1/nm. (c) Fourier transformation of the High-resolution HAADF STEM*

*image of 6R TaS$_2$ (inset, Scale bar, 5 nm) shows low-frequency spots (yellow arrows), suggesting a commensurate phase in the crystal.*

Among TMDs, TaS$_2$ has a unique place due to the exciting interplay between CDWs and superconductivity. Both 1T and 2H forms of TaS$_2$ show CDWs,[29, 30] which are in direct competition with superconducting pairing. To probe the CDW in 6R TaS$_2$, we have performed further electrical measurements above room temperature (Figure 3a). These measurements showed two closely spaced resistance transitions at 320 K and 305 K. Similar transitions (350 K and 180 K) were also observed in IT TaS$_2$ and were attributed to the transition from the incommensurate (IC) CDW phase to nearly commensurate (NC) CDW phase and from nearly commensurate to the commensurate CDW (CCDW) phase respectively[31]. In the low-temperature commensurate phase, the Ta atoms of the 1T layers displace to form a commensurate superstructure with 13 Ta atoms arranged in the shape of Star of David as depicted in the Schematic (Figure 3a). The NC structure also possesses such atomic arrangement, but in distant domains separated by a dis-commensuration network [31]. To further probe the CDWs in 6R TaS$_2$, we performed systemic temperature-dependent TEM studies from room temperature to 103 K. Figure 3b shows the selected area electron diffraction pattern from 6R TaS$_2$ at room temperature (293 K). The additional diffraction spots (yellow circles in Figure 3b inset) show the presence of CDW. Similar spots were also observed in 1T TaS$_2$ but only below 180 K[31] suggesting the observed 305 K transition in 6R TaS$_2$ is to a commensurate phase. We have also performed a temperature-dependent cryogenic electron diffraction study of the samples down to 103 K and observed no appearance of additional spots in the diffraction pattern of the sample (Figure S10). A room temperature HRSTEM image of 6R TaS$_2$ crystal shows a hexagonal arrangement of Ta atoms (Figure 3c inset). Fourier transformation of the image shows the appearance of six singlet spots in the first order **q** positions (marked with yellow arrows, Figure 3c). The presence of singlet spots further confirms the existence of the commensurate phase in 6R TaS$_2$ at room temperature rather than NC phase, where the spots in first-order *q* positions appear as triplets[32]. Raman spectra of bulk 6R sample also confirm the presence of commensurate structure in the 1T planes at room temperature (Figure S6). Further high-temperature electron diffraction or scanning tunnelling microscopy studies are required to fully confirm the nature of transition observed at 320 K. However, based on the 1T TaS$_2$ CDWs, we attribute the transition at 320 K to the IC CDW phase to NC CDW phase.

One very interesting aspect of layered materials is the ability to separate them into single layer forms by chemical or physical exfoliation processes. To understand the effect of the reduced dimensionality on the superconductivity of 6R TaS$_2$, we have exfoliated the as-prepared 6R TaS$_2$ single crystal by Li intercalation (Supporting information, Figure S11) followed by liquid-phase exfoliation using ultrasonication. The Li intercalated TaS$_2$ was also found to be superconducting but with an enhanced $T_c$ of 3.0 K (Figure S12). The liquid exfoliated layers in 6R TaS$_2$ were then restacked (well separated and electronically decoupled) to obtain random stacking of layers in 6R TaS$_2$ (Supporting information). It is to be noted that 6R TaS$_2$ consists of alternating 1H and 1T planes that separate from each other during exfoliation. Magnetic and transport measurements on the restacked 6R TaS$_2$ showed an increased superconducting $T_c$ of 3.6 K (Figure S13a-c). We observe a huge shift in CDW transition temperature in restacked 6R phase from NC to commensurate phase to 250 K from 320 K in bulk 6R TaS$_2$ (Figure S13d), switching closer to the transition temperature observed in the 1T TaS$_2$. Superconducting $T_c$ and CDW $T_c$ of restacked TaS$_2$ samples closely resemble that of 1H and 1T layers, respectively. This indicates that 1T and 1H layers in 6R TaS$_2$ are separated into 2H and 1T phases during exfoliation and restacking. We have also studied the superconducting $T_c$ of mechanically exfoliated thin layers (~1 nm) of 6R TaS$_2$ and found its $T_c$ similar to the bulk 6R TaS$_2$ (Supporting information, Figure S14).

To explain the underlying mechanisms responsible for the emergence of superconductivity and CDWs in 6R TaS$_2$, we have performed first-principles calculations of its electronic and phononic properties, as well as the electron-phonon coupling and the resulting superconducting state (Supporting information). The calculations were performed using density functional (perturbation) theory (DF(P)T), as implemented in the ABINIT package[33]. We started our investigation from the 1T and 1H monolayers (MLs), and the 1T-1H bilayer (BL), to establish a thorough bottom-up comparison between the elementary TMD phases and heterogeneous phases like the 1T-1H BL. 6R TaS$_2$ is composed of three such 1T-1H BLs arranged with rhombohedral stacking (see Figure 1c).

We focus first on the electronic properties of the different TaS$_2$ structures around the Fermi level, directly relevant for their superconducting and CDW properties. The Fermi surface of the 1T-1H BL (Figure S15f) is essentially a combination of the individual Fermi surfaces of the two MLs (Figure S15d,e), with similar Fermi surface shapes and corresponding Fermi velocities ($v_F$).

Nevertheless, there are some interesting effects of interlayer interactions. Firstly, an avoided crossing occurs along the Γ-M direction, between the 1T- and 1H-based Fermi sheets (the former positioned around M, the latter around Γ and K). Secondly, there is a spin-orbit coupling (SOC)-induced splitting in the 1T-based sheet, due to the lack of inversion symmetry in the BL originating from the 1H layer. The Fermi surface of the bulk 6R phase (Figure 4a) clearly is the 3D counterpart of the Fermi surface of the BL (Figure S15f), with nearly identical Fermi sheets and Fermi velocity distribution.

The close similarity between the 6R phase and the constituent 1T-1H BLs persists in our DF(P)T results on the phonon density of states (PHDOS) and $\alpha^2F$, the Eliashberg spectral function of the electron-phonon coupling (Figure 4c,d and Figure S16c,f). Moreover, we found the resulting electron-phonon coupling constant of the 6R phase ($\lambda$=1.02, Figure 4d) to be very close to that of the 1T-1H BL ($\lambda$ = 1.04, Figure S16f), but also to $\lambda$ of the 1H ML ($\lambda$ = 1.07, Figure S16d). This provides clear proof that 6R $TaS_2$ indeed hosts superconductivity, in agreement with our magnetic and transport measurements. Moreover, these results indicate the superconducting phase of 6R $TaS_2$ to be driven by the 1H planes.

As our experimental results have revealed a CDW state akin to that of bulk 1T $TaS_2$ to occur in 6R $TaS_2$, we set out to explore its microscopic origins through first-principles calculations. Both the 1H and 1T ML display CDW-type instabilities in their phonon dispersions - as shown in Figure 4h,i - which have been resolved by lowering the broadening factor of the Fermi-Dirac smearing function for the electronic occupation (see Supporting information for the details). The phonon dispersion of the 1H ML (Figure 4h) shows an instability around M, leading to a simple integer lattice reconstruction (into an $n \times n$ supercell, with $n$ a natural number ranging from 2 to 8 according to our DFPT result). On the other hand, the phonon dispersion of the 1T ML (Figure 4i) shows more complex behaviour, as it relates to the $\sqrt{13} \times \sqrt{13}$ lattice reconstruction known as Star-of-David. Through an analogous calculation on the 1T-1H BL, shown in Figure 4j, we found that both CDW types coexist in this system, albeit spatially separated in the respective layers.

Altogether, our first-principles results indicate that the 1T layers in 6R $TaS_2$ will be insulating at temperatures well below the CCDW-$T_c$ (305 K in 6R $TaS_2$, Figure 3a), owing to the Mott state. Thus, the occurrence of ML H-type $TaS_2$, surrounded by insulating 1T layers, can explain the increase in superconducting $T_c$ in our bulk 6R $TaS_2$ sample, far exceeding the $T_c$ of bulk 2H TaS$_2$. Such enhancements were also noted for 2H TaS$_2$ samples where the superconducting 2H layers were decoupled by intercalation[34].

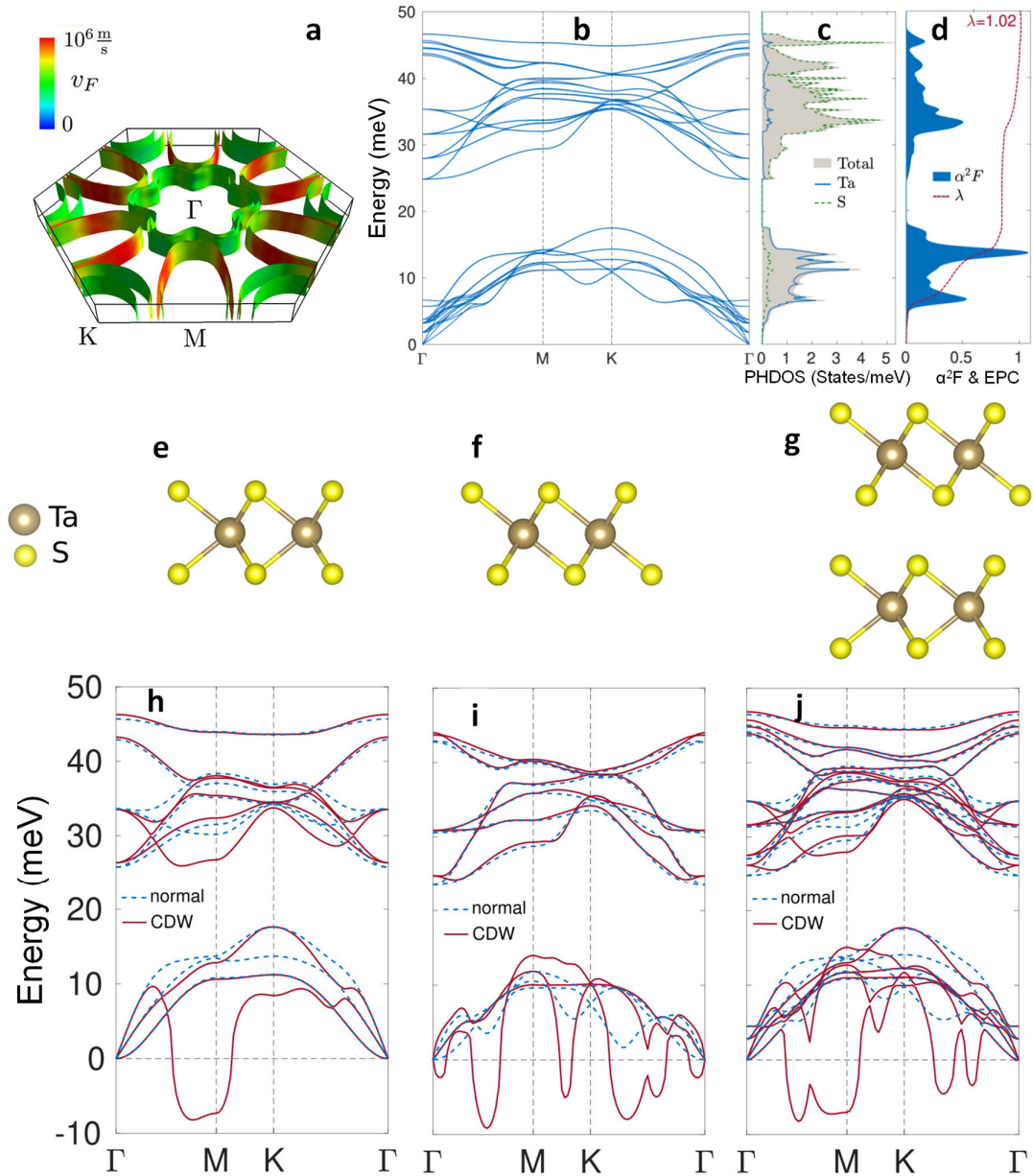

*Figure 4. First-principles calculations of electronic and phononic properties of 6R TaS$_2$. (a) Fermi surface of 6R TaS$_2$, where the colours indicate the Fermi velocities. (b) Phonon band structure of 6R TaS$_2$, (c) the corresponding total and atom-resolved phonon DOS (PHDOS), and (d) the Eliashberg function α$^2$F and resulting electron-phonon coupling constant λ. Crystal structures of (e) ML 1H TaS$_2$, (f) ML 1T TaS$_2$, and (g) T-H heterobilayer. The corresponding phonon band structures in both the normal (dashed blue lines) and charge density wave (CDW) regimes (solid red lines) of (h) ML 1H TaS$_2$, (i) ML 1T TaS$_2$, and (j) T-H heterobilayer.*

In conclusion, we have reported the phase transition of 1T TaS$_2$ into 6R TaS$_2$ when heated at 800 °C. The as prepared 6R TaS$_2$ shows the coexistence of superconductivity with a $T_c$ of 2.6 K and CDW transitions at 320 K and 305 K. Our TEM analysis shows the presence of a commensurate CDW phase at room temperature. Exfoliation and random restacking of layers in 6R TaS$_2$ enhance the superconducting $T_c$ to 3.6 K while decreasing the NC to CCDW transition to 250 K. This superconducting $T_c$ and CDW transition temperatures closely resembles that of monolayer 1H and few-layer 1T TaS$_2$ samples, respectively, indicating separation of 1T and 1H sheets in 6R TaS$_2$. Our first-principles calculations show the coexistence of the main electronic, vibrational and electron-phonon coupling properties of individual T and H layers in mixed T-H layers in 6R TaS$_2$. This suggests that the origin of superconductivity in 6R TaS$_2$ lies in the 1H layers of the crystal, separated from one another by insulating 1T layers in the CCDW state, scarcely interfering with the superconducting state. This alternating layered structure makes this material a true 2D superconductor in bulk form and opens a plethora of intriguing questions related to Josephson physics, THz radiation, etc. Further work is needed to understand the dynamics of reported phase transition in TaS$_2$. This could enable the controlled synthesis of different polytypes of layered chalcogenide materials.

ACKNOWLEDGMENTS

This work was supported by the Royal Society, the Engineering and Physical Sciences Research Council (EP/N005082/1), and European Research Council (contract 679689). The authors acknowledge the use of the facilities at the Henry Royce Institute and associated support services. J.B. is a postdoctoral fellow of Research Foundation-Flanders (FWO-Vlaanderen). Computational resources were provided by the VSC (Flemish Supercomputer Center), funded by the FWO and the Flemish Government department EWI. This work was also performed under a transnational access provision funded by the European Union under the Horizon 2020 programme within a contract for Integrating Activities for Advanced Communities No 823717 – ESTEEM3

# Supporting Information

**Materials and Methods**

**Preparation of 6R TaS$_2$ and *in* situ XRD characterisation:** Single crystalline 1T and 2H TaS$_2$ crystals were obtained from HQ graphene and used as such. A piece of 1T TaS$_2$ crystal (dimension approx. 2 mm x 2 mm) was cleaved from the single crystal and placed on a heating stage of Rigaku diffractometer with a graphite dome. The crystal was heated under vacuum (approx. $10^{-3}$ bar) at 800 °C for 5 h to convert it into 6R form.

Phase transition in the 1T TaS$_2$ crystals was probed by in-situ X-ray diffraction studies using a Rigaku Smartlab XRD system (Cu K$\alpha$, $\lambda$=0.15406 nm). A small piece (4 x 4 mm) of 1T TaS$_2$ sample was placed on the heating stage of the XRD instrument. The position and height of the sample were carefully aligned so that the sample was properly exposed to the X-ray beam. A K$\beta$ filter was used to remove extra peaks from the Cu K$\beta$ wavelength. The heating stage containing the sample was fitted with a graphite dome and a vacuum of $10^{-3}$ mbar was applied. The sample was heated at 30 °C/min and spectra were collected at regular temperature intervals. The sample was stabilised at each temperature before XRD data was collected.

Powder XRD data of the samples were collected using a D-8 Discover advanced XRD system (Cu K$\alpha$, $\lambda$=0.15406 nm). The samples were ground in a mortar before performing the powder XRD.

Raman spectra was measured with HORIBA Raman spectrometer (XploRA PLUS) with a laser excitation of 532 nm (spot size ~1 μm, laser power of 0.125 mW and spectrometer grating of 1200 grooves per millimeter) for measuring the Raman spectra of the samples.

**Electron microscopy:** A state-of-the-art double corrected Thermo Fisher Titan QU-Ant-EM transmission electron microscope was used for TEM imaging. High-resolution HAADF-STEM images were acquired at 300 kV using a convergence semi-angle $\alpha$ of 21 mrad, 50 pA probe current and Cs tuned close to 0 μm. To minimize distortions during long scanning in STEM mode, a time series technique was employed.

In-situ cooling experiments were performed on a Thermo Fisher Osiris transmission electron microscope operated at 200 kV. Temperature experiments were performed using a Gatan 636 double-tilt liquid nitrogen cooling holder.

Cross-sectional TEM foils were prepared using FIB from bulk 1T and 6R $TaS_2$ samples using the lift out method in a Thermo Fisher dual beam FIB/SEM instrument. A protective platinum layer using electron beam assisted deposition has been used. An ion beam of 2 kV/0.2 nA was employed to achieve the final thinning and to minimize defects generated during high voltage FIB thinning on both sides of the sample.

TEM data were processed with custom-made scripts based on the open-source python libraries *Hyperspy*[1] and *Pixstem*[2].

**Magnetic measurements:** A Quantum Design Magnetic Property Measurement System (MPMS-3) was used to measure the temperature- and field-dependent magnetization of the samples. In the zero-field cooling (ZFC) mode, the samples were initially cooled to 1.8 K in zero applied field, then a desired external field *H* was applied and the magnetisation *M* was measured as a function of increasing temperature, T (typically 1.8–20 K). The field-cooling (FC) part of an *M (T)* curve was obtained on cooling the sample to 1.8 K in the same *H*.

The magnetisation measurements were performed in two different geometries. For in-plane measurements, samples were mounted on a quartz rod with GE varnish keeping the c-axis perpendicular to the direction of the magnetic field. Out of plane measurements were done by placing the crystal at the bottom of a gelatine capsule and mounting it in the magnetometer using a straw sample holder in such a way that the c-axis of the crystal was parallel to the magnetic field direction. Since both sample holders had a negligible magnetic response, the magnetic data was used without any background correction.

**Electrical transport measurements:** For electrical transport measurements, we have fabricated $TaS_2$ devices in a linear four-probe geometry. Four contacts were made with silver paint on a 2 × 4 mm rectangle sample of a $TaS_2$ crystal. The samples were loaded on an Electrical Transport Option (ETO) sample holder and sample rod in MPMS-3. The ETO is capable of measuring differential resistance as a function of current or voltage. This is achieved by applying a small AC excitation on top of a DC offset bias. The AC response is measured and used to calculate the differential resistance. The differential resistance is a direct measure of the first derivative of the IV curve at a given DC bias.

**Preparation of exfoliated 6R $TaS_2$:** Exfoliated samples of 6R $TaS_2$ were prepared by lithium intercalation of $TaS_2$ followed by exfoliation. Lithium intercalation of the crystals was done inside a glove box. In a typical procedure, a small piece of 6R $TaS_2$ crystal (approx. 5 mg) was immersed in 0.1 mL of 1.6 M solution of n-butyl lithium in a glass vial. The vial was rested for 48 hours before washing the crystal with hexane three times. The lithium intercalated crystal was dried and used for liquid-phase exfoliation of 6R $TaS_2$. Liquid phase exfoliation was performed by sonicating lithium intercalated 6R $TaS_2$ in water for 1 h to exfoliate into flakes. Such exfoliated samples were unstable in water and prone to oxidation if stored in water for longer than 24 hours. The dispersion was quickly centrifuged at 12000 rpm, followed by washing 3 times with water and 1 time with ethanol. The washed sample was dried under a vacuum for magnetic measurements.

**Electrical measurements using mechanically exfoliated 6R $TaS_2$:** The $TaS_2$ layers were exfoliated from bulk crystals onto $SiO_2$/Si substrates in an argon-filled glovebox with levels of $O_2$ and $H_2O$ below 0.5 ppm. Thin layers of $TaS_2$ (~ 1nm) were identified by using optical contrast under an optical microscope. Since mechanically exfoliated thin layers of $TaS_2$ are prone to oxidation, to prevent any exposure of $TaS_2$ to air during measurements, thin hexagonal boron nitride crystal was used as the capping layer and few-layer graphite strips as contact electrodes. They were aligned and transferred onto the monolayer in the Ar-filled glovebox using the standard dry transfer process with polypropylene carbonate (PPC) coated polydimethylsiloxane (PDMS) films as stamps. Later, 1D metal contacts to the graphene strips were fabricated by standard methods as reported in our previous work[3].

The critical temperature of thin $TaS_2$ (outlined by the red dotted line in Figure S14a) was measured using exfoliated strips of graphene as the contact electrodes (outlined by the black dotted lines in Figure S14a). The current-voltage (*I-V*) characteristics of the device were measured by applying the small AC excitation current of 10 nA and sweeping the DC bias current (Figure S14b). While the electric current was applied across the flake of 6R-$TaS_2$ through the graphene electrodes, voltage is measured on the graphene electrode that is only in contact with the few-layer 6R-$TaS_2$. This is to avoid any contribution from multilayer $TaS_2$ that is in contact with the other electrode (see Figure S14a). Dips in the measured differential resistance (*dV/dI*) are signatures of superconductivity. Since the contact between the graphene electrode and $TaS_2$ is not transparent, *dV/dI* increases near zero bias. By carefully examining the temperature dependence of *dV/dI* as a function of DC biasing current (Figure S14c), $T_c$ is determined to be ~ 2.4 K, where the non-linear *I-V* characteristics due to superconductivity disappear.

**First-principles calculations:** Our density functional theory (DFT) calculations make use of the Perdew-Burke-Ernzerhof (PBE) functional implemented within the ABINIT code[4]. We included spin-orbit coupling (SOC) using fully relativistic Goedecker pseudopotentials[5,6]. Here, Ta-$5d^36s^2$ and S-$3s^23p^4$ states were included as valence electrons, together with an energy cutoff of 50 Ha for the plane-wave basis. Van der Waals interactions were included via the vdw-DFT-D3 method introduced by Grimme,[7] based on the Becke-Jonhson formalism[8]. All crystal structures were relaxed so that forces on each atom were below 1 meV/Å. To simulate the 2D structures (ML 1T and 1H and the 1T-1H BL), 25 Å of vacuum was included in the unit cells. In all density functional perturbation theory (DFPT) calculations, a 24 × 24 × 1 k-point grid and a 12 × 12 × 1 q-point grid were employed. SOC was omitted in the DFPT calculation on 6R $TaS_2$, due to computational restrictions posed by the size of its unit cell.

To identify CDWs, we have compared DFPT results obtained with higher (0.01 Ha) and lower (0.0025 Ha) broadening factors of the Fermi-Dirac smearing function for the electronic occupations. The case with lower broadening resolves the phonon instabilities corresponding to the CDW (Figure 5). To evaluate the electron-phonon coupling (EPC) responsible for the superconducting state, we have used Migdal-Eliashberg theory[9], based on the EPC matrix elements obtained through the DFPT calculations.

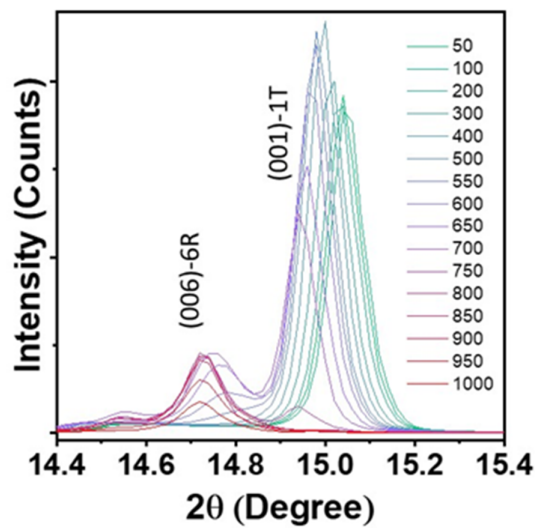

**Figure S1. In-situ XRD showing the transition of a single crystalline 1T TaS$_2$ into 6R phase.** Phase conversion of the 1T TaS$_2$ samples was measured in-situ on a heated XRD stage at regular temperature intervals as shown in the figure. At temperatures below 500 °C, we only observed a monotonous downshifting of (001) peak associated with the thermal expansion of the crystalline c axis. Above 600 °C we noticed the appearance of a new peak at lower 2θ (14.78°) compared to the original peak of the 1T phase at 15.05°.

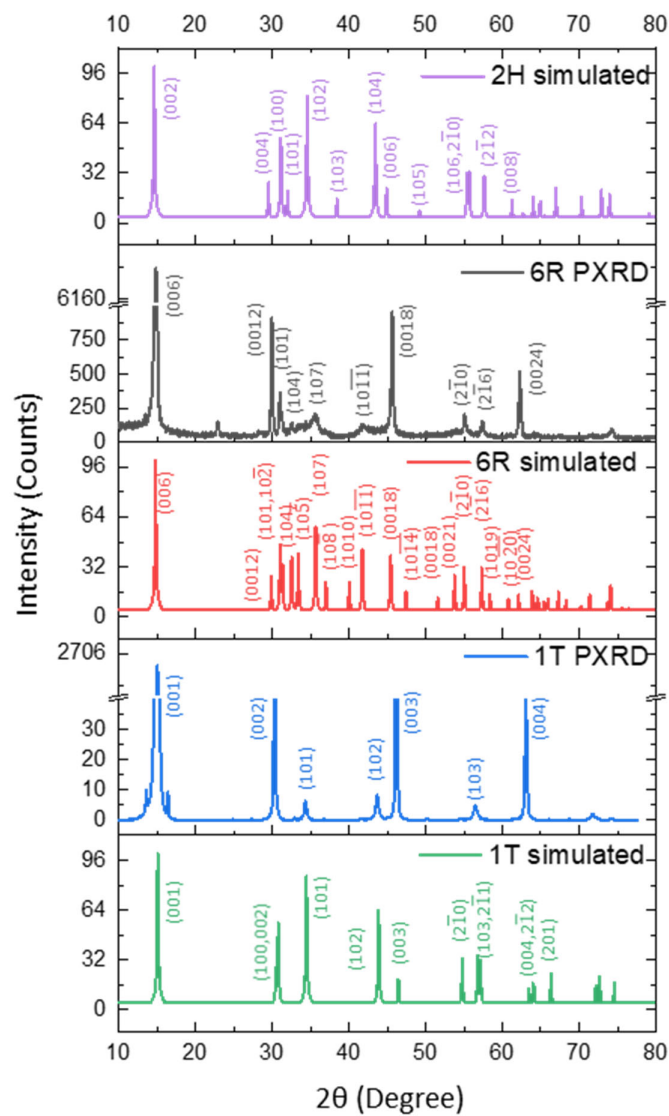

**Figure S2. Power XRD of 6R and 1T TaS$_2$.** Comparison of powder XRD pattern of 6R TaS$_2$ with that of 1T, and simulated PXRD patterns of 1T, 2H and 6R phases. The PXRD pattern of 6R TaS$_2$ shows peaks matching with (104), (107), (10$\overline{11}$) peaks in the simulated 6R phase, which are not present in 1T or 2H TaS$_2$.

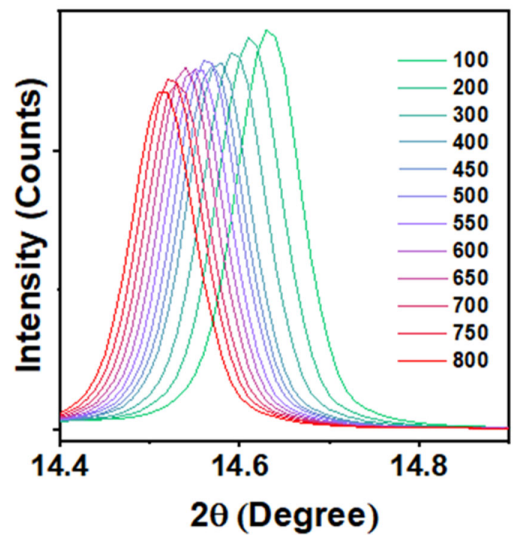

**Figure S3. In-situ XRD during annealing of 2H TaS$_2$.** Temperature-dependent XRD of 2H TaS$_2$ heated up to 800 °C showing the monotonous shift in the (002) peak. 2H being the most stable phase among the TaS$_2$ polytypes, we do not see any transition into the 6R phase up to 800 °C. The downshifting of the peak is due to the thermal expansion of the crystal along the c-axis. After cooling, the peak reverts to its original position.

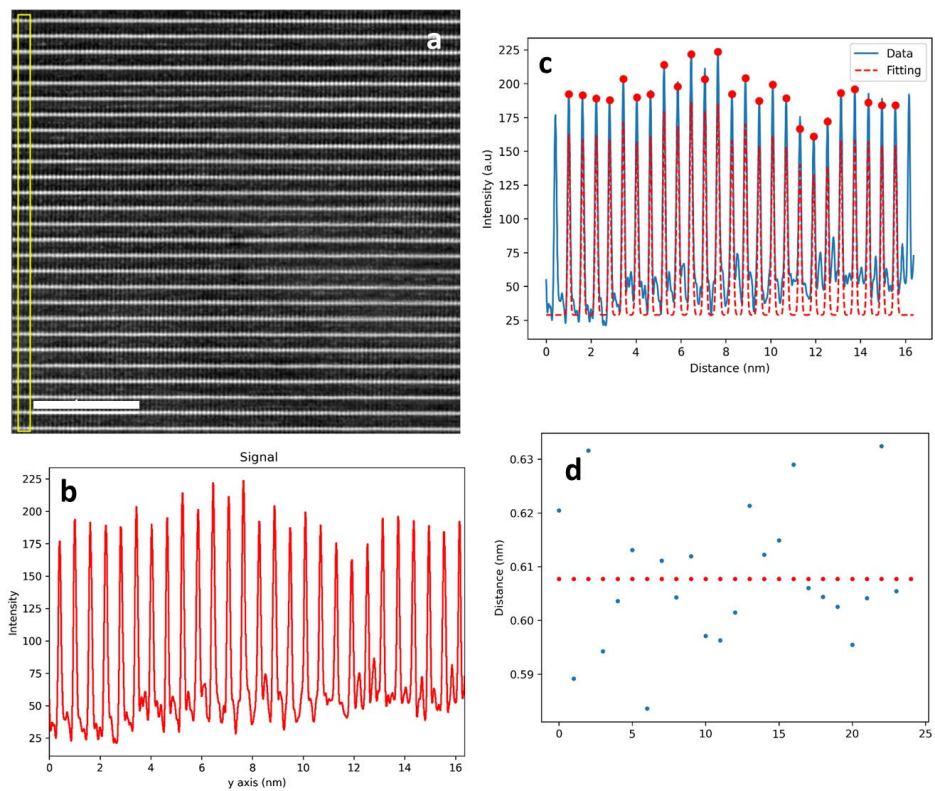

**Figure S4. Analysis of HRSTEM data.** (a) HRSTEM image of 6R TaS$_2$ cross-section showing an example rectangle section along which intensity profiles were estimated. (Scale bar, 4 nm). The average from five such regions was plotted in (b). The red dotted lines in figure (c) correspond to the fitted Gaussian function for every intensity profile in (b). The red circles represent the centre of the Gaussians. (d) d-spacing (blue dots) and it's mean (red dotted line) along (00l) TaS$_2$ direction are plotted according to the position of the centre Gaussians (red circles in Figure c).

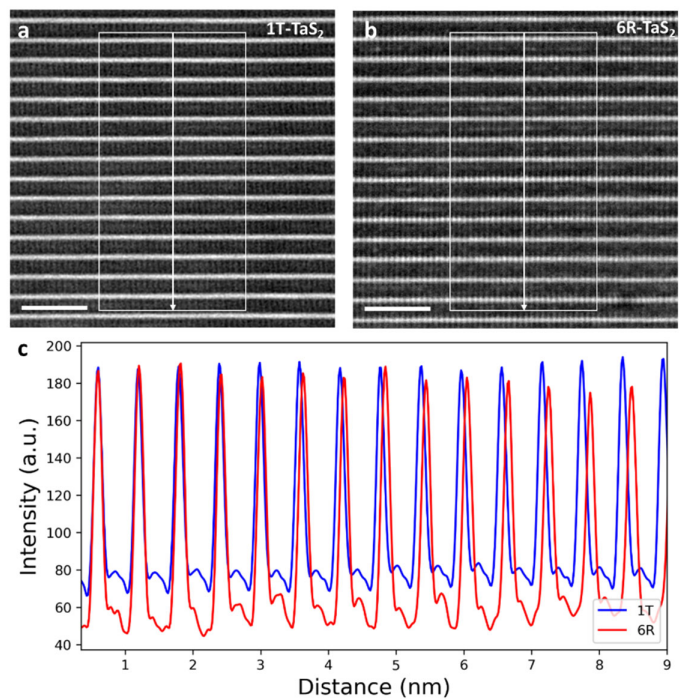

**Figure S5. Lattice expansion in 6R TaS$_2$.** High-resolution STEM images of the pristine 1T (a) and after the heat treatment at 800 °C in vacuum (b). Scale bar, 2nm (c) Comparison of the intensity profiles along (0001) direction taken from the rectangular regions in a, b shows an increase in interlayer spacing after heating.

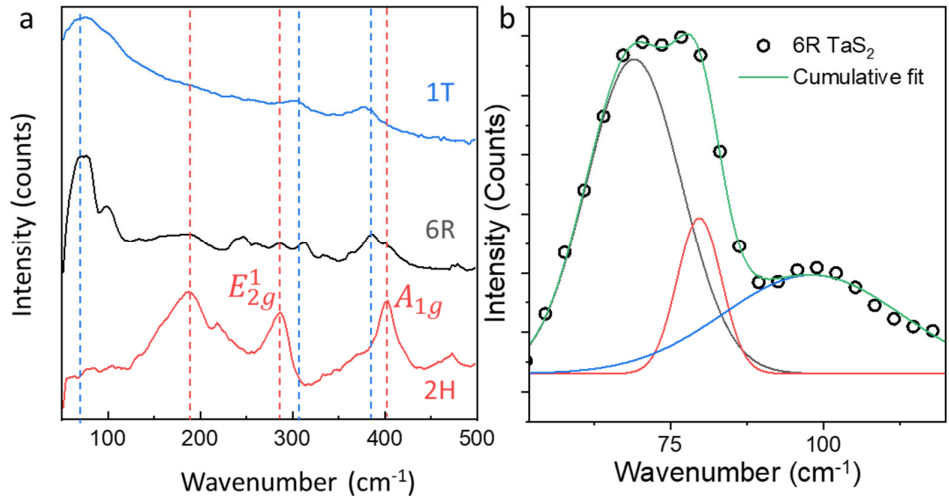

**Figure S6. Raman spectra of 6R, 1T, and 2H TaS$_2$.** (a) Comparison of Raman spectra between 1T, 2H and 6R TaS$_2$ samples showing the presence of features from both 1T and 2H phase in the 6R sample. The 2H sample shows typical $E_{2g}^1$ and $A_{1g}$ Raman modes at 280 and 400 cm$^{-1}$ respectively[10]. The peak at 180 cm$^{-1}$ is observed due to second order scattering[11]. All of these peaks appear in the 6R sample. (Red dotted lines) On the other hand, the peaks observed for bulk 1T TaS$_2$ at 80, 303 and 384 cm$^{-1}$ corresponding to the nearly commensurate structure at room temperature[12] are also present in the 6R sample (Blue dotted lines) but with additional features. Interestingly, the low frequency peak of 6R sample becomes sharper and well defined compared to the 1T phase with the appearance of a peak at 243 cm$^{-1}$, signifying the presence of commensurate structure[13]. (b) Enlarged view of the low frequency modes in 6R TaS$_2$ showing deconvoluted peaks at 70, 80 and 98 cm$^{-1}$, these phonon bands originate from folding of the Brillouin zone signifying CCDW phase transition in the 1T layers of 6R TaS$_2$ at room temperature[14].

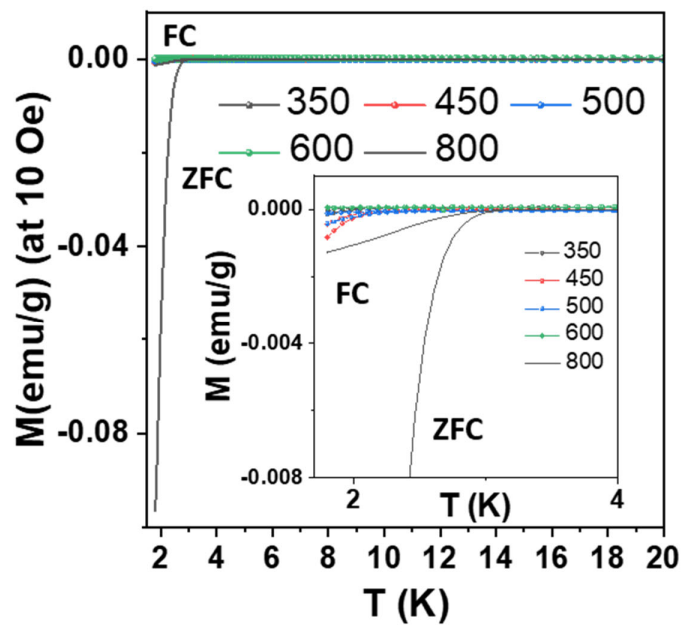

**Figure S7. 1T to 6R transition seen in magnetisation measurements.** Temperature dependence of ZFC and FC magnetisation, M, for single-crystalline 1T TaS$_2$ heated at different temperatures under vacuum at 10 Oe magnetic field. Inset: zoomed in magnetic transition.

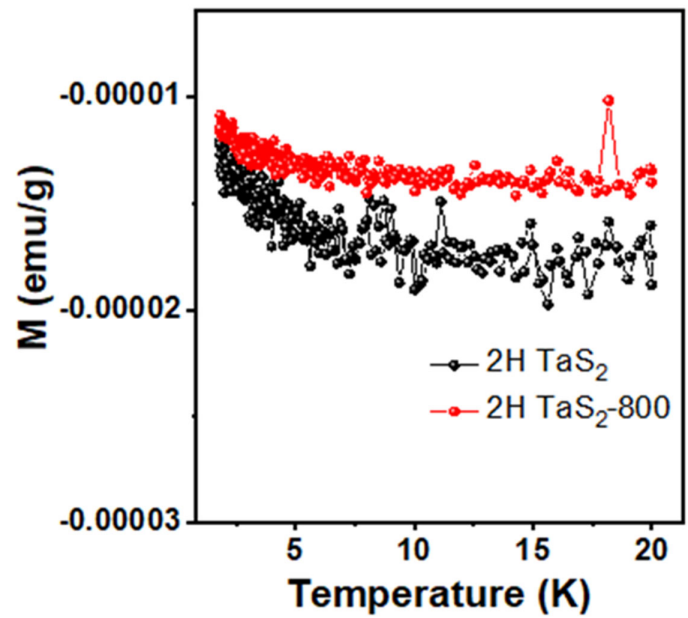

**Figure S8. Low temperature magnetisation of 2H TaS$_2$ crystal before and after heating.** ZFC and FC temperature-dependent magnetisation of 2H TaS$_2$ and 2H TaS$_2$ heated at 800 °C under vacuum, showing no sign of superconductivity down to 1.8 K.

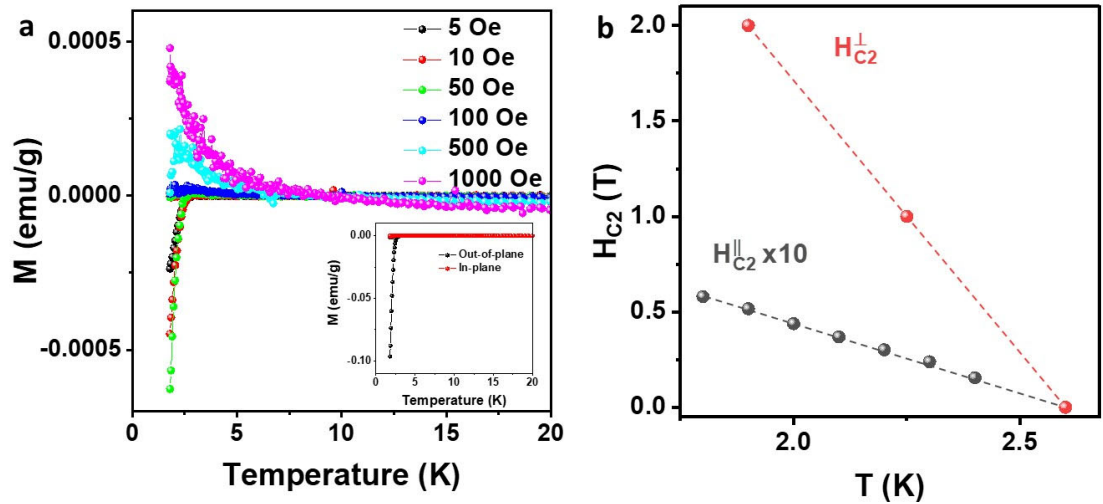

**Figure S9. Anisotropic superconductivity in 6R TaS$_2$.** a) ZFC and FC temperature-dependent magnetisation of 6R TaS$_2$ with its c-axis perpendicular to the magnetic field. Inset: Comparison of in-plane ($ab\|H$) and out-of-plane ($c\|H$) magnetisation measurement carried out at 10 Oe. b) $H_{C2}$ as a function of temperature for in-plane (*c*-axis perpendicular to the applied field, $H_{C2}^{\perp}$, red spheres) and out-of-plane (*c*-axis parallel to the applied field, $H_{C2}^{\|}$, black spheres) superconductivity. For better visibility $H_{C2}^{\|}$ has been multiplied by 10.

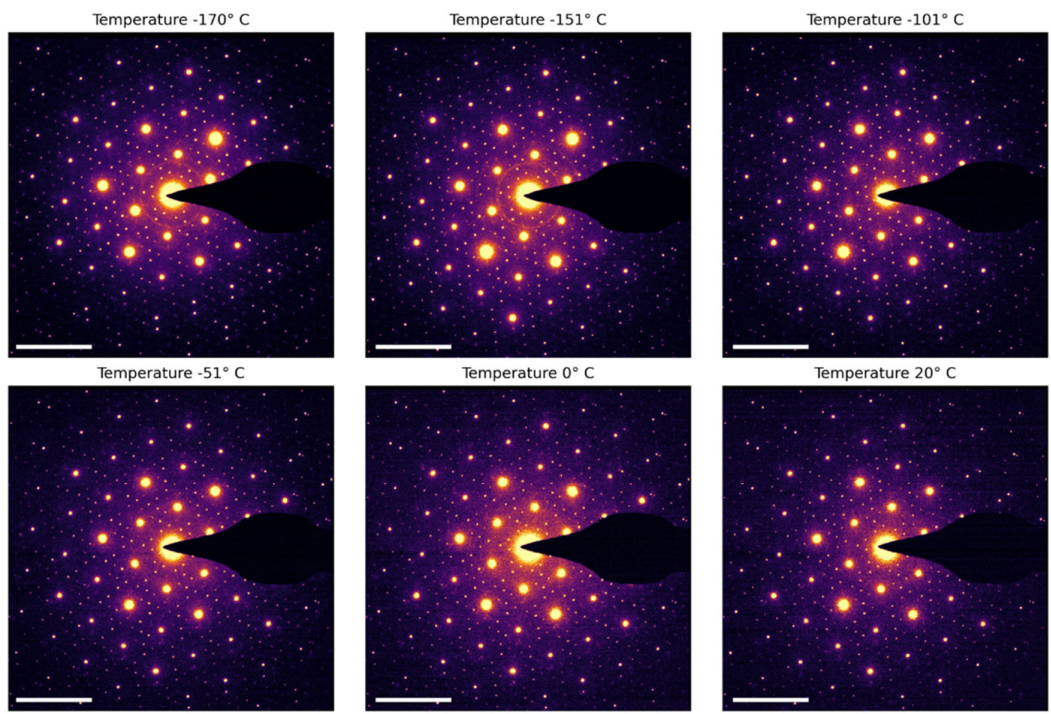

**Figure S10. Low temperature electron diffraction.** Selected area electron diffraction patterns acquired during the in-situ cooling experiment in the range (-170 °C to 20 °C). No additional diffractions spots were noticed while cooling the sample. Scale bar, 6 1/nm.

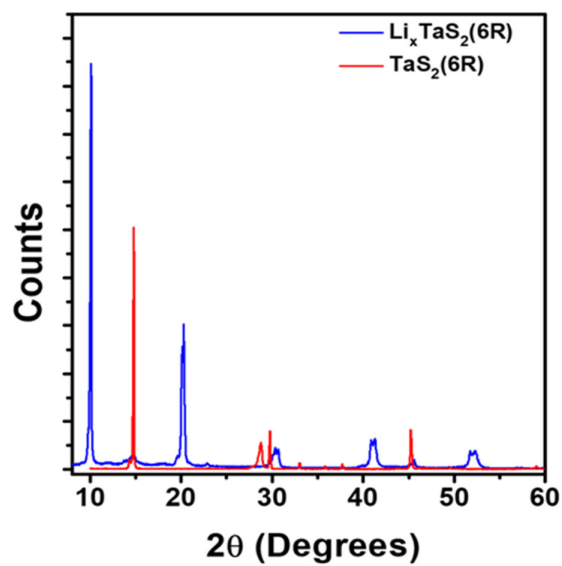

**Figure S11. Lithiation of 6R TaS₂. XRD pattern of 6R TaS₂ and lithiated 6R TaS₂ measured on single crystalline samples.** The crystals are oriented on their ab plane; hence only 00l peaks appear in the XRD. From the XRD spectra, the low angle peak (006) of the 6R TaS₂ shifts from 14.8° to 10° after lithiation, indicating the intercalation of lithium into the interlayer spaces. The interlayer spacing for the intercalated sample was 0.873 nm compared to 6R TaS₂ with a d spacing of 0.598 nm. It is to be noted that the lithiated sample was prone to oxidation in air and hence was stored in a glove box under an inert atmosphere.

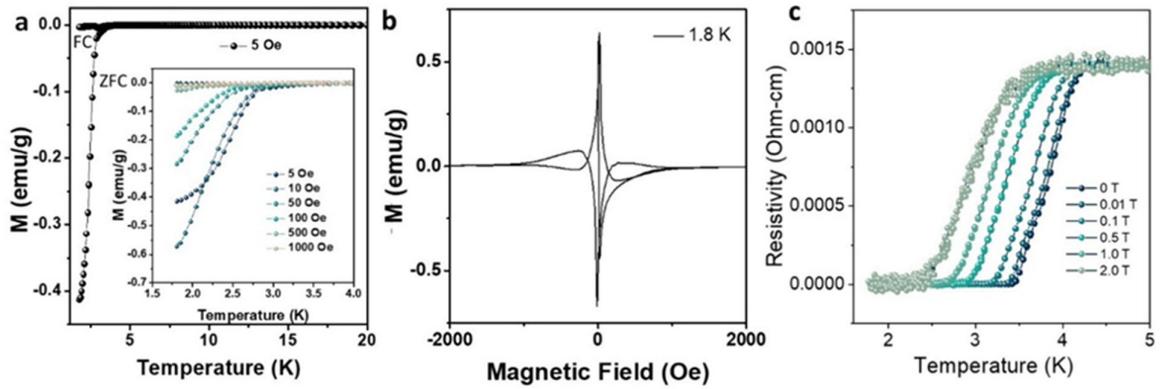

**Figure S12. Superconductivity in lithiated 6R TaS$_2$.** (a) Temperature dependence of magnetisation of single-crystalline 6R TaS$_2$ intercalated with lithium under the magnetic field of 5 Oe, with zero-field-cooling curve and field-cooling curve. Inset shows field dependence of transition temperature of lithiated 6R TaS$_2$. We observed an increase in transition temperature from 2.6 K in as-prepared 6R TaS$_2$ to 3.0 K when lithium was introduced in the system. (b) Magnetic hysteresis of lithium intercalated 6R TaS$_2$ at 1.8 K, showing type 2 superconducting nature. (c) Field dependence of electrical resistance of lithiated 6R TaS$_2$ showing superconducting transition at ~ 3K.

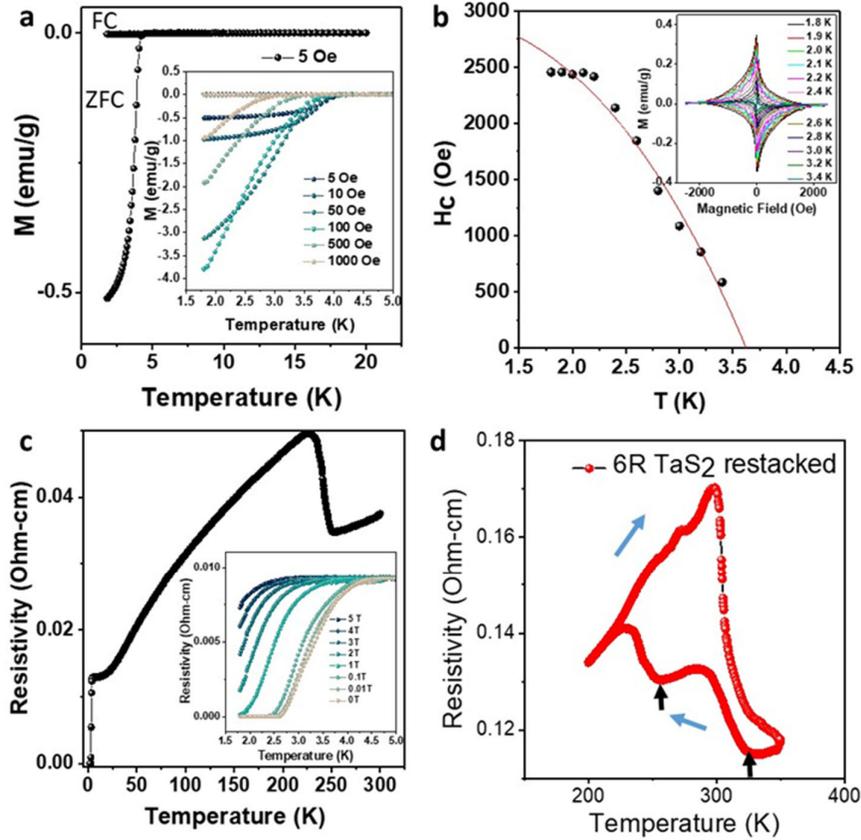

**Figure S13. Superconductivity in restacked 6R TaS$_2$.** (a) Temperature dependence of magnetisation of restacked 6R TaS$_2$ under the magnetic field of 5 Oe, with zero-field-cooling curve and field-cooling curve. Inset shows field dependence of transition temperature of restacked 6R TaS$_2$. (b) Temperature dependence of upper critical field Hc$_2$. The upper critical field was calculated from the divergence point in the M-H hysteresis curves (inset). The red line represents the best least squares fit of the equation $H_{C2}(T) = H_{C2}[1 - (T/T_C)^{(1+\alpha)}]$, where $H_{C2}$ and $\alpha$ are the fitting variables[15]. (c) Temperature dependence of electrical resistivity of restacked 6R TaS$_2$ at H = 0 T. inset: Field dependence of electrical resistance of restacked 6R TaS$_2$ (d) Temperature-dependence of electrical resistivity of restacked 6R TaS$_2$ showing CDW transitions at 250 K and 320 K from nearly commensurate to commensurate and incommensurate to nearly commensurate phases respectively (marked by black arrows). Blue arrows denote temperature sweep direction. It is to be noted that the restacked samples were not stable in air for more than 24 hours; hence we performed all measurements immediately after preparing the samples.

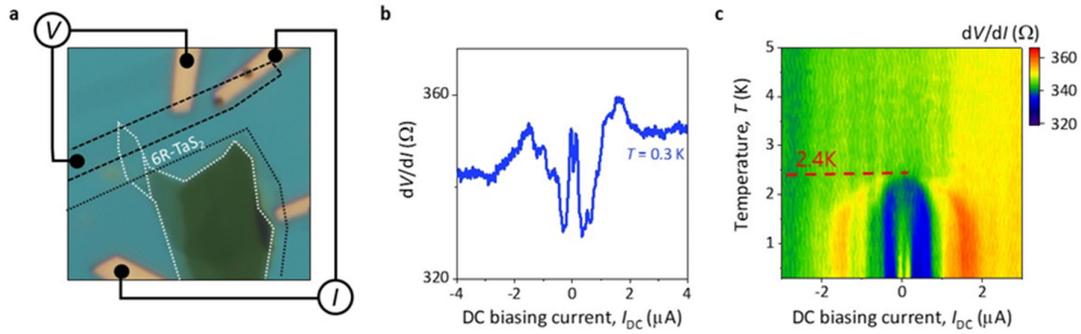

**Figure S14. Electron transport measurements of mechanically exfoliated TaS$_2$.** (a) Optical micrograph of few-layer 6R-TaS$_2$ device with schematics of the wiring used for measurements. The black dotted line outlines the thin graphite electrodes and the white dotted line outlines the TaS$_2$ flakes. The entire TaS$_2$ and graphite electrodes were capped by a thin layer of hexagonal boron nitride crystal. (b) Measured differential resistance ($dV/dI$) as a function of DC biasing current ($I_{DC}$) at temperature $T$= 0.3 K. (c) $dV/dI$ as a function of $I_{DC}$ and temperature $T$. The red dotted line indicates the superconducting transition temperature, above which the non-linear $I$-$V$ characteristics disappear.

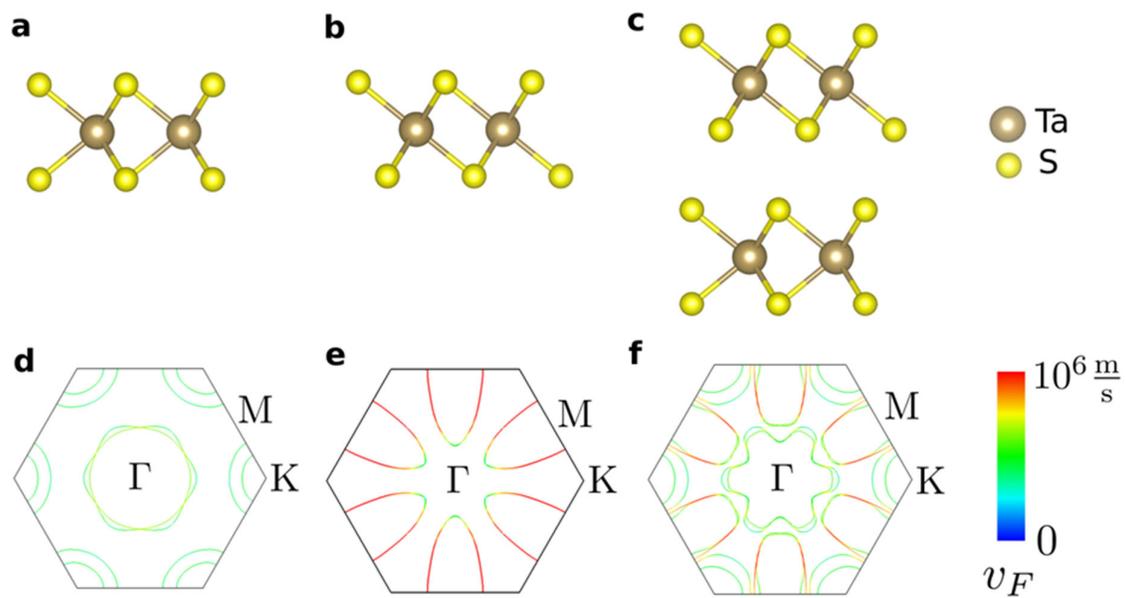

**Figure S15. Fermi surfaces of monolayer and bilayer TaS$_2$.** Crystal structure of (a) ML 1H TaS$_2$, (b) ML 1T TaS$_2$, and (c) BL 1T-1H TaS$_2$. Fermi surface of (d) ML 1H TaS$_2$, (e) ML 1T TaS$_2$ and (f) BL 1T-1H TaS$_2$, where the colours indicate the Fermi velocities.

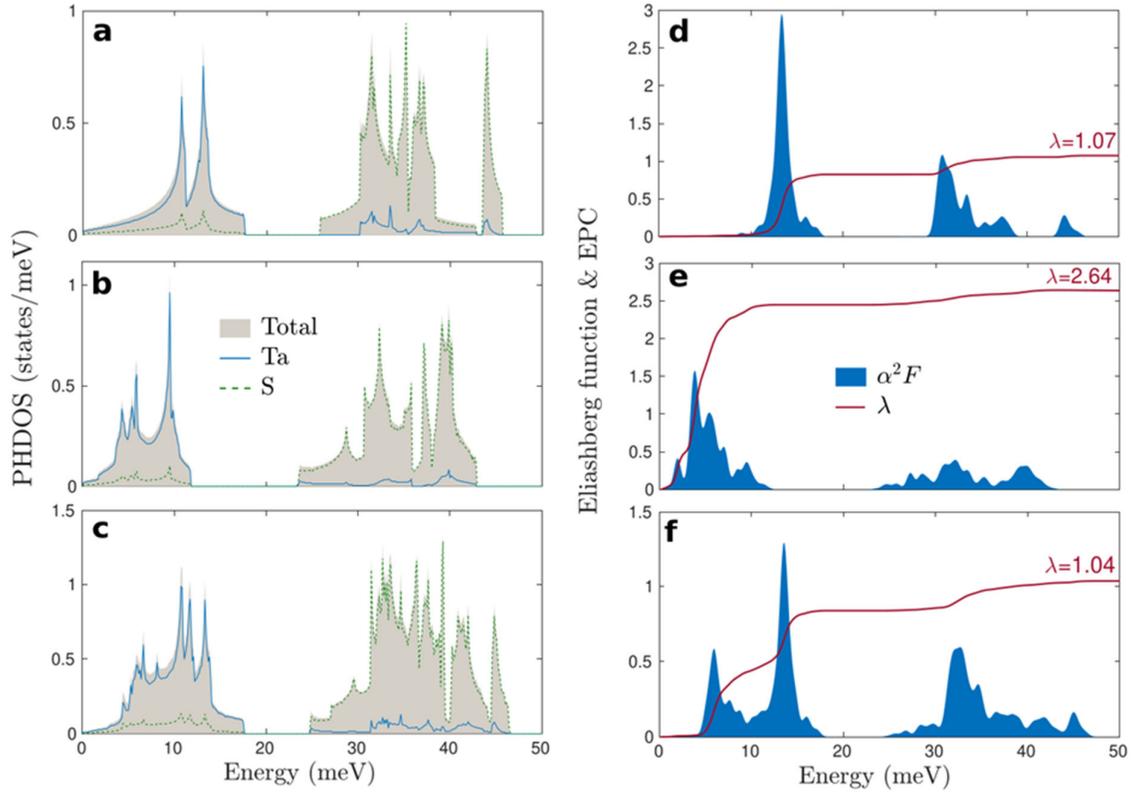

**Figure S16. Phonon density of states and electron-phonon coupling in monolayer and bilayer TaS$_2$.** Total and atom-resolved phonon density of states (PHDOS) of (a) ML 1H TaS$_2$, (b) ML 1T TaS$_2$ and (c) BL 1T-1H TaS$_2$, and Eliashberg function ($\alpha^2$F) and electron-phonon coupling (EPC, λ) of (d) ML 1H TaS$_2$, (e) ML 1T TaS$_2$ and (f) BL 1T-1H TaS$_2$.